\documentclass[aps,twocolumn,prd,showpacs,nofootinbib]{revtex4}
\usepackage{amsmath}
\usepackage{graphicx}
\usepackage{dcolumn}
\usepackage{bm}
\usepackage{amssymb}
\usepackage{latexsym}

\def\setC{\mathbb{C}}

\def\setR{\mathbb{R}}

\newcommand{\si}[1]{{\scriptscriptstyle{#1}}}
\newcommand{\GN}{G_{_{\rm N}}}
\newcommand{\dd}{\mathrm{d}}
\newcommand{\ee}{\mathrm{e}}

\newcommand{\ie}{\textsl{i.e.~}}
\newcommand{\eg}{\textsl{e.g.~}}

\newcommand{\mP}{M_{_\mathrm{Pl}}}
\newcommand{\GReCO}{${\cal G}\setR\varepsilon\setC{\cal O}$}

\def\spose#1{\hbox to 0pt{#1\hss}}
\def\lta{\mathrel{\spose{\lower 3pt\hbox{$\mathchar"218$}}
     \raise 2.0pt\hbox{$\mathchar"13C$}}}
\def\gta{\mathrel{\spose{\lower 3pt\hbox{$\mathchar"218$}}
     \raise 2.0pt\hbox{$\mathchar"13E$}}}

\newcommand{\rhoeff}{\rho_\mathrm{eff}}
\newcommand{\peff}{p_\mathrm{eff}}

\begin{document}

\title{Regular cosmological bouncing solutions in \\ low energy
effective action from string theories.}

\author{J. C. Fabris}
\email{fabris@cce.ufes.br}
\affiliation{Departamento de F\'{\i}sica, Universidade Federal do
Esp\'{\i}rito Santo, 29060-900, Vit\'oria, Esp\'{\i}rito Santo,
Brazil}

\author{R. G. Furtado}
\email{furtado@cce.ufes.br}
\affiliation{Departamento de
F\'{\i}sica, Universidade Federal do Esp\'{\i}rito Santo,
29060-900, Vit\'oria, Esp\'{\i}rito Santo, Brazil}

\author{Patrick Peter}
\email{peter@iap.fr}
\affiliation{Institut d'Astrophysique de
Paris, \GReCO, FRE 2435-CNRS, 98bis boulevard Arago, 75014 Paris,
France}

\author{N. Pinto-Neto}
\email{nelsonpn@cbpf.br}

\affiliation{Lafex - Centro Brasileiro de Pesquisas F\'{\i}sicas --
CBPF, \\ rua Xavier Sigaud, 150, Urca, CEP22290-180, Rio de Janeiro,
Brazil}

\date{15 May 2003}

\begin{abstract}
The possibility of obtaining singularity free cosmological solutions
in four dimensional effective actions motivated by string theory is
investigated. In these effective actions, in addition to the
Einstein-Hilbert term, the dilatonic and the axionic fields are also
considered as well as terms coming from the Ramond-Ramond sector. A
radiation fluid is coupled to the field equations, which appears as a
consequence of the Maxwellian terms in the Ramond-Ramond
sector. Singularity free bouncing solutions in which the dilaton is
finite and strictly positive are obtained for models with flat or
negative curvature spatial sections when the dilatonic coupling
constant is such that $\omega < - 3/2$, which may appear in the so
called $F$ theory in 12 dimensions. These bouncing phases are smoothly
connected to the radiation dominated expansion phase of the standard
cosmological model, and the asymptotic pasts correspond to very large
flat spacetimes.
\end{abstract}

 \maketitle

\section{Introduction}

Superstring is the most promising candidate to describe a unified
theory of all interactions, gravity included. There are five
consistent superstring theories in 10 dimensions, which are connected
among themselves through duality transformations. To each superstring
theory, there is a corresponding supergravity theory in 10
dimensions. All of them can be obtained from the 11 dimensional
supergravity theory. This indicates that those superstring theories
are different manifestations of a unique 11 dimensional framework,
that has been named
$M$ theory~\cite{polchinski,green,kiritsis}. Moreover, the superstring
type-IIB can be recast in a more geometrical form in a 12 dimensional
model, suggesting that perhaps a yet more fundamental framework may
exist in 12 dimensions, which has been called $F$ theory~\cite{pope}.

The physical properties of superstring theories become relevant at
energy scales comparable with the Planck scale. This renders very
improbable that superstring phenomenology may be tested in the near
future in some laboratory experiment (see, however, Ref.~\cite{randal}
in which the Planck mass is lowered to TeV scale by accounting for
large extra dimensions). According to the hot big bang scenario,
however, energy scales even as high as the usual Planck scale
($\mP\sim 10^{19}$GeV) may have been reached in the very early
universe. Hence, for the moment, cosmology seems to be the most
natural arena where the consequences of superstring theories may be
tested. The pre-big bang paradigm~\cite{PBB} was one of the first
ideas to implement superstring theories in this framework. Some relics
of a cosmological string phase may also be identified~\cite{bdp},
opening perhaps the possibility of testing superstring models.
Furthermore, superstring theories open the possibility that some
typical drawbacks of the standard cosmological model, such as the
existence of an initial singularity, may be solved in the context of
superstring cosmological models. The goal of the present paper is to
show that, under certain conditions, it is possible to obtain
completely regular bouncing cosmological models in the context of
effective actions constructed from superstring theories (not
involving, in particular, negative energies~\cite{ppnpn2}), for which,
moreover, the dilaton is strictly positive (nonvanishing) at all times
and never diverges.

String cosmology is based on the low energy limit of string or
superstring theories. In the most general case of the supersymmetric
string theory, there are two sectors, related with the choice of
periodic or antiperiodic boundary conditions on the spinor fields,
namely the Ramond and Neveu-Schwarz (NS)
sectors~\cite{polchinski,green}. Since fermions can be either left or
right moving, this leads to four possible combinations of these
sectors. The bosonic fields arise both from the NS-NS and
Ramond-Ramond (RR) sectors. The NS-NS sector provides the
Einstein-Hilbert term, as well as a three-form, called the axionic
field, and the dilaton. The latter is directly related with the string
perturbative expansion parameter and takes the form of a
Brans-Dicke-like scalar field, nonminimally coupled to both the
Einstein-Hilbert and the axionic fields. In the RR sector, $p$ forms
appear, which are minimally coupled to the dilaton field. The
dimensions of the $p$ forms depend on the specific superstring theory
which is under consideration. In cosmological applications, we are
interested in scalar fields that emerge from these different $p$
forms. They differ by the way they couple with the dilatonic field and
between themselves.

Out of the many different possibilities stemming from string
theory, one can construct in general an effective action suitable
for cosmological applications with two main features: scalar
fields coming from the NS-NS sector and nonminimally coupled to
the dilaton, and scalar fields from the RR sector which are
minimally coupled to the dilaton. All the possibilities are not
exhausted by these two frameworks, but they summarize the general
aspects of what has been proposed in the literature as far as
effective actions coming from string theory are concerned. One
can also obtain phenomenological matter fields by averaging on
some components of those original $p$ forms.

This brief description explicits the great richness of the string
effective action procedure, which implies a large variety of possible
cosmological models. Notice that these effective actions exhibit great
similarities with those that can be obtained from multidimensional and
supergravity theories. To select one cosmological model that could be
a candidate to describe the physical world, two possible prescriptions
are: either the cosmological model is completely regular, with no
curvature or expansion parameter singularity, or it is compatible with
observation; ultimately, both criteria should be satisfied. In the
present work we will concentrate on the first. The second criterion,
which presents some specific challenges, will be treated in the
future~\cite{future}.

The search for singularity free cosmology in string theories is not a
new subject~\cite{lidsey,vasquez,picco,kirill,branden}. The string
action at tree level does not lead in general to singularity free
cosmological solutions, at least when the strict string case ($\omega
= - 1$, $\omega$ being the dilatonic coupling parameter) is
considered. The pre-big bang model~\cite{PBB}, which is an example of
a string cosmology, requires the introduction of nonlinear curvature
terms in order to achieve a smooth transition from a curvature growing
phase to a curvature decreasing phase. If large negative values of the
dilatonic coupling parameter $\omega$ are allowed, it is possible, in
some cases, to obtain completely regular models, including in the
dilatonic sector~\cite{kirill}. This may be achieved mainly in models
with spatial sections with negative curvature.

Here, it will be shown that regular cosmological models may also be
obtained if a radiation fluid is coupled to the string action at the
tree level. Such a radiation fluid can have a fundamental motivation,
for example, in the case of the superstring type IIB theory, where a
5-form appears in the RR sector. Truncation and dimensional reduction
of this 5-form lead to a Maxwell term in four dimensions with the
desired features~\cite{fabris}. Hence, the model to be studied here is
totally based on superstring theories.  The string motivated
phenomenological term included under the form of a radiation fluid
makes it possible to connect smoothly such string cosmological models
to the radiation phase of the standard cosmological model before
nucleosynthesis.

In Ref.~\cite{picco}, models motivated by string theory similarly
including a radiation fluid have been studied, restricted to flat
spatial sections and $\omega > -3/2$.  In such cases, bouncing
solutions have been obtained only for $\omega < -4/3$. Furthermore,
for these solutions, the dilaton vanishes in the infinite past,
raising doubts on the validity of the tree level action in such a
region. In the present paper, the curvature of the spatial sections
and the value of $\omega$ are kept arbitrary. New bouncing regular
solutions are then obtained, for which, as mentioned above, the
dilaton remains finite and nonvanishing at all times. When $\omega >
-3/2$, which includes the strict string case ($\omega = -1$), the
solutions can only be bouncing provided the spatial sections have
negative curvature and if the dilaton is always negative, which is not
consistent with the higher dimensional framework of stringlike
theories, and implies a repulsive gravity.  We shall henceforth
disregard such solutions. When $\omega < -3/2$, the bouncing solutions
are obtained for models with flat or negative curvature, and the
dilaton is strictly positive at all times.

In the following section, we derive effective string motivated actions
in four dimensions. In particular, we show how to obtain an effective
string action in four dimensions with $\omega < -3/2$ in the context
of the so-called $F$ theory in twelve dimensions. In Sec. III we
derive nonsingular cosmological solutions from these effective
theories, which are thoroughly discussed in Sec. IV from the point of
view of violation of energy conditions. We end up with the conclusions
in Sec. V.

\section{The effective action}\label{sec:omega}

Our analysis is based on the following effective action at tree level:
\begin{eqnarray}
L = \sqrt{-\tilde g}\ee^{-\tilde\sigma}\biggl(\tilde R -
\omega\tilde\sigma_{;\si{A}}\tilde\sigma^{;\si{A}} -
\frac{1}{12}H_{\si{ABC}}H^{\si{ABC}}\biggr)\nonumber \\ - \sqrt{-
\tilde g}\biggr(\frac{1}{2}\xi_{;\si{A}}\xi^{;\si{A}} +
\frac{1}{240}F_{\si{ABCDE}}F^{\si{ABCDE}}\biggr)\label{lagrange}
\end{eqnarray}
where $\tilde\sigma$ is the dilatonic field, $H_{\si{ABC}}$ is the
axionic field, and $\omega$ is the dilatonic coupling constant.  The
two last terms come from the RR sector of superstring type IIB. The
tildes indicate that all quantities are considered in a
$D$-dimensional spacetime, $D = 10$ in the pure superstring context.

The dilatonic coupling constant is $\omega = - 1$ for usual
superstring theory. However, this may not necessarily be the case for
some ten dimensional theories stemming from a more fundamental one in
higher dimensions. In some specific situations, the value of $\omega$
can be found to be even less than $-3/2$.  As an example, the
superstring type IIB action may be reformulated in 12 dimensions, in
the context of the so-called $F$ theory. A low energy limit of the
$F$ theory has been studied by~\cite{pope}, where an action in 12
dimensions has been established, which was shown to lead to the low
energy limit of the superstring type IIB in 10 dimensions through
truncation and dimensional reduction. Let us consider this twelve
dimensional action, given by~\cite{pope}
\begin{eqnarray}
L_{12} = \sqrt{-\tilde g}\biggl(\tilde R -
\frac{1}{2}\Psi_{;\si{A}}\Psi^{;\si{A}} -
\frac{1}{48}\ee^{a\Psi}F_{\si{ABCD}}F^{\si{ABCD}} \nonumber \\
-\frac{1}{240}\ee^{b\Psi}G_{\si{ABCDE}}G^{\si{ABCDE}} + \lambda
B_4{\scriptstyle \wedge} \dd A_3{\scriptstyle \wedge} \dd A_3\biggr),
\label{f12}
\end{eqnarray}
with $a^2 = - 1/5$ and $b^2 = - 4/5$, $\lambda$ being a coupling
parameter for the Chern-Simons type term involving the potentials of
the five and four-forms. Writing the metric as
\begin{equation}
\dd s_{12}^2 = g_{\mu\nu}\dd x^\mu \dd x^\nu - \ee^{2\beta}\dd x_i \dd
x^i ,
\end{equation}
with Greek indices $\mu, \nu$ running from 0 to 9 and Latin indices
$i\in [10,11]$, and setting the five-form equal to zero, we obtain the
following Lagrangian
\begin{eqnarray}
L_{10} & = & \sqrt{-g}\ee^{2\beta}\biggr(R +
2\beta_{;\rho}\beta^{;\rho} - \frac{1}{2}\Psi_{;\rho}\Psi^{;\rho}
\nonumber \\ &- &\frac{1}{12}\ee^{a\Psi - 2\beta}
F_{\mu\nu\lambda}F^{\mu\nu\lambda} - \frac{1}{8}\ee^{a\Psi -
4\beta}F_{\mu\nu}F^{\mu\nu}\biggl),
\end{eqnarray}
where we have retained just the two and three-forms coming from the
four-form in the original action. The term originating the three-form
was made purely imaginary in 12 dimensions. Choosing $\Psi = 2\beta/a$,
and defining $\phi = \ee^{2\beta}$, one ends up with the following
action in 10 dimensions:
\begin{equation}
L_{10} = \sqrt{-g}\biggl[\phi\biggl(R +
3\frac{\phi_{;\rho}\phi^{;\rho}}{\phi^2} -
\frac{1}{12}F_{\mu\nu\lambda}F^{\mu\nu\lambda}\biggl) -
\frac{1}{8}F_{\mu\nu}F^{\mu\nu}\biggr]\label{f10}.
\end{equation}
One can see that, in this case, we obtain an action with $\omega = -
3$ together with a Maxwell term (which generates the radiation
fluid). This is a remarkable example of how an effective string action
with $\omega \neq -1$ (in this case, $\omega = -3 < -3/2$) can be
realized. That is why we will maintain the value of $\omega$ in
Eq.~(\ref{lagrange}) arbitrary in what follows, unless otherwise
specified.

The $D$-dimensional metric is written as
\begin{equation}
\dd s^2 = g_{\mu\nu}\dd x^\mu \dd x^\nu -
\ee^{2\beta}\delta_{ij}\dd x^i \dd x^j,
\end{equation}
where $g_{\mu\nu}$ is the four dimensional metric, $\ee^\beta$ is the
scale factor of the $d = D - 4$ dimensional internal space which we
suppose to be homogeneous and flat. For now on, we will consider a
static internal space. This is not obligatory in some of the cases to
be analyzed latter, but such a restriction considerably simplifies the
unified presentation of many different cases allowed by the action
given by Eq.~(\ref{lagrange}).

Dimensional reduction and isotropization of the Maxwellian term [which
may come from the RR sector, or as described in the passage from
Eq.~(\ref{f12}) to Eq.~(\ref{f10})], lead to the following effective
action in four dimensions:
\begin{equation}
\label{l1} {\it L} = \sqrt{-g}\biggl[\phi\biggl(R -
\omega\frac{\phi_{;\rho}\phi^{;\rho}}{\phi^2} -
\frac{\Psi_{;\rho}\Psi^{;\rho}}{\phi^{2}}\biggr) -
\frac{1}{2}\xi_{;\rho}\xi^{;\rho}\biggr] + L_{\rm r},
\end{equation}
where from now on Greek indices run from 0 to 3.  In this action,
$\phi = \ee^{-\tilde\sigma}$ is the dilaton, the field $\Psi$ comes
from the axionic term, and is thus called the axion, while $L_{\rm r}$
represents an ordinary radiation fluid term, which can be obtained
from the five-form existing in the RR sector, as was stressed
above. We shall also call $\xi$ the RR-scalar as it originates from
the same sector.

The Lagrangian described by Eq.~(\ref{l1}) may cover theories others
than pure string theory. For instance, one can consider a more general
coupling between the dilatonic and axionic fields, \ie of the type
$\phi^{n}\Psi_{;\rho}\Psi^{;\rho}$, with $n$ a new parameter. This less
restrictive coupling contains the string case if one sets $n = - 1$,
but general multidimensional theories and supergravity theories in
higher dimensions are examples where the parameter $n$ may take
different values. As was discussed in Refs.~\cite{picco,fabris}, the
final results depend very weakly on the parameter $n$ and one may thus
expect that this generalized effective action will give results that
are essentially similar to those obtained in the pure string
case. Hence, the results presented below are of a quite general
nature, and should not be understood as statements restricted to
string cosmology only, even though they were derived in this specific
framework.

{}From Eq.~(\ref{l1}), we obtain the field equations
\begin{eqnarray}
R_{\mu\nu} - \frac{1}{2}g_{\mu\nu}R &=& \frac{8\pi}{\phi} T_{\mu\nu} +
\frac{\omega}{\phi^2}\biggr(\phi_{;\mu}\phi_{;\nu} -
\frac{1}{2}g_{\mu\nu}\phi_{;\rho}\phi^{;\rho}\biggl) \nonumber \\ & &
+\frac{1}{\phi}\biggr(\phi_{;\mu\nu} - g_{\mu\nu}\Box\phi\biggl)
\nonumber\\ & &+\frac{1}{\phi^2}\biggr(\Psi_{;\mu}\Psi_{;\nu} -
\frac{1}{2}g_{\mu\nu}\Psi_{;\rho} \Psi^{;\rho}\biggl) \nonumber \\ & &
+ \frac{1}{\phi}\biggr(\xi_{;\mu}\xi_{;\nu} -
\frac{1}{2}g_{\mu\nu}\xi_{;\rho}\xi^{;\rho}\biggl),
\label{eq:Einstein}
\end{eqnarray}
for the Einstein part,
\begin{equation}
\Box\phi + \frac{2}{3 + 2\omega}\phi^{-1}\Psi_{;\rho}\Psi^{;\rho} +
\frac{1}{3 + 2\omega}\xi_{;\rho}\xi^{;\rho} = \frac{8\pi T}{3 +
2\omega},
\end{equation}
with $T\equiv T^\mu_{\ \ \mu}$ the trace of the stress-energy tensor,
for the dilaton $\phi$, while we get
\begin{equation}
\Box\Psi - \Psi_{;\rho}\frac{\phi^{;\rho}}{\phi}=0,
\end{equation}
to describe the dynamics of the axion $\Psi$, and finally
\begin{eqnarray}
\Box\xi &=& 0,\\ {T^{\mu\nu}}_{;\mu} &=& 0,
\end{eqnarray}
for the RR-scalar $\xi$ and the radiation fluid respectively. These
equations we now implement in a cosmological context.

\section{Cosmological solutions}

Introducing the Friedman-Robertson-Walker metric
\begin{equation}
\dd s^2 = \dd t^2 - a^2(t)\left[\frac{\dd r^2}{1 - kr^2} +
r^2\left(\dd\theta^2 + \sin^2\theta \dd\phi^2\right)\right],
\end{equation}
$k$ being the normalized curvature of the maximally symmetric spatial
sections ($k = 0, \pm 1$), and assuming the fields now depend only on
time, the field equations derived above reduce to the following
equations of motion:
\begin{equation}
\label{em1} 3\biggr(\frac{\dot a}{a}\biggl)^2 + 3\frac{k}{a^2}
= 8\pi\frac{\rho}{\phi} +
\frac{\omega}{2}\biggr(\frac{\dot\phi}{\phi}\biggl)^2 - 3\frac{\dot
a}{a} \frac{\dot\phi}{\phi} + \frac{\dot\Psi^2}{2\phi^{2}} +
\frac{\dot\xi^2}{2\phi},
\end{equation}
which is the generalization of the Friedman equation, and
\begin{eqnarray}
\label{em2}
\ddot\phi + 3\frac{\dot a}{a}\dot\phi + \frac{2}{(3 +
2\omega)}\frac{\dot\Psi^2}{\phi} + \frac{\dot\xi^2}{(3 + 2\omega)} &=&
\frac{8\pi(\rho - 3p)}{(3 + 2\omega)},\nonumber \\ \\ \label{em3}
\ddot\Psi + 3 \frac{\dot a}{a}\dot\Psi - \dot\Psi\frac{\dot\phi}{\phi}
&=& 0,\\
\label{em4}
\ddot\xi + 3\frac{\dot a}{a}\dot\xi &=& 0,\\
\label{em5} \dot\rho + 3\frac{\dot a}{a}(\rho + p) &=& 0.
\end{eqnarray}
In these expressions, $\rho$ is the energy density and $p$ is the
pressure of some perfect fluid which obeys, for the sake of
generality, a barotropic equation of state, $p = \lambda\rho$, with
$\lambda$ an arbitrary constant. In what follows, we will specialize
this fluid to the case we are interested in, namely, radiation, for
which $\lambda = \lambda_\mathrm{r}=1/3$. A dot stands for a
derivative with respect to the cosmic time $t$.

Equations (\ref{em3}), (\ref{em4}) and (\ref{em5}) admit the first
integrals
\begin{equation}
\dot\Psi = \frac{A\phi}{a^3}, \quad \dot\xi = \frac{B}{a^3}, \quad
\rho = D a^{- 3(1 + \lambda)}, \label{solPsi}
\end{equation}
where $A$, $B$ and $D$ are integration constants. According to the
string motivated action discussed above, let us now specialize the
equations for the radiation fluid case and accordingly set $\lambda =
1/3$. For this specific case, Eq.~(\ref{em2}) simplifies to
\begin{equation}
\label{scalar} \ddot\phi + 3\frac{\dot a}{a}\dot\phi + \frac{2}{(3
+ 2\omega)}\frac{A^2}{a^6}\phi + \frac{B^2}{(3 + 2 \omega)a^6} = 0,
\end{equation}
which can be solved in the following way. It is convenient to define a
new timelike coordinate $\theta$ given by the relation
\begin{equation}
\dd t = a^3 \dd\theta.
\label{parameter}
\end{equation}
In terms of this new coordinate, Eq.~(\ref{scalar}) reads
\begin{equation}
\label{scalar'} \phi'' + \frac{2A^2}{(3 + 2\omega)}\phi +
\frac{B^2}{(3 + 2 \omega)} = 0,
\end{equation}
where primes denote differentiations with respect to $\theta$.
Similarly, Eq.~(\ref{em1}), when expressed in terms of $\theta$, reads
\begin{equation} \left({a'\over a^3}\right)^2 +k={M\over a^2 \phi} +
{\omega \over 6} {\phi'^2 \over a^4 \phi^2} - {a'\phi'\over a^5 \phi}
+{1\over 6 a^4} \left( A^2 + {B^2\over
\phi}\right),\label{aprim}\end{equation} in which use has been made of
Eq.~(\ref{solPsi}),
and we have set $M = 8\pi D/3$, which is dimensionless in the
radiation case.

Equation (\ref{aprim}) may be recast in a very convenient form through
the redefinition
\begin{equation}
a = \phi^{-1/2}b,
\end{equation}
which implies to change to the so-called Einstein frame. This yields
\begin{equation}
\left({b'\over b}\right)^2 + ( kb^2-M) {b^2\over\phi^2} = {1\over
6}\left( A^2 +{B^2\over \phi} + {3+2\omega\over 2}
{\phi'^2\over\phi^2}\right), \label{bprim}
\end{equation}
whose solution we next investigate. Notice, however, that we want to
keep considering the Jordan frame as the physical frame; the conformal
transformation above is introduced only for technical reasons. The
solutions of Eqs.~(\ref{scalar'}) and (\ref{em1}), with the
redefinition made above for the scale factor, depend on the sign of
the term $3 + 2\omega$ and on the presence of the RR scalar field. We
will consider each case separately. For simplicity, we will call $3 +
2\omega > 0$ ($< 0$) as the normal (anomalous) case, and $\xi =$ const
($\xi \neq$ const) as the axionic (RR) case.

In the special case for which $\omega$ assumes the critical value
$\omega = -3/2$, the dilatonic field is not a physical degree of
freedom since it may be eliminated by means of a conformal
transformation $g_{\mu\nu} = \ee^{-\phi}\tilde g_{\mu\nu}$: it is a
mere artifact of a metric redefinition. For $\omega < - 3/2$, the
scalar field in the Einstein frame, obtained from the Jordan frame by
a conformal transformation, does not preserve energy conditions as it
appears with a sign for its kinetic term which is opposite to the
usual situation, leading to a negative energy contribution [see the
last term in Eq.~(\ref{bprim})]. Such a negative energy field can
provide the necessary compensation with the usually positive energy
density contributions in order to allow bouncing solutions in general
relativity~\cite{ppnpn2,ppnpn1}.  For $\omega > - 3/2$, the last term
in Eq.~(\ref{bprim}) appears with the ordinary sign, implying a
positive energy contribution.

In what follows, the quantities $\phi_0$ and $a_0$ are constants of
integration subject to the constraints indicated in each case.

\subsection{Normal axionic case}\label{sec:na}

In this first case for which $\xi$ is constant [\ie $B=0$ in
Eq.~(\ref{solPsi})] and $\omega > -3/2$, the solution of
Eq.~(\ref{scalar'}) is given by
\begin{equation}
\phi(\theta) = \phi_0\sin(\alpha\theta),\label{phiNA}
\end{equation}
where
\begin{equation}
\alpha = \sqrt{\frac{2A^2}{3 + 2\omega}}
\end{equation}
and we have chosen $\phi (0)=0$.

Plugging this solution into Eq.~(\ref{bprim}) yields
\begin{equation}
\phi_0^2 \sin^2(\alpha\theta) b'^2 = (C^2 + M b^2 - k b^4 ) b^2,
\label{bNA}
\end{equation}
where
\begin{equation}
C^2 = \frac{1}{6}A^2\phi_0^2.\label{C}
\end{equation}
We are seeking regular bouncing solutions for which the scale factor
is bounded from below but can grow arbitrarily large, while $\phi$ is
nonvanishing and finite. This means that the function $b$ should also
grow indefinitely on both sides of the bounce. As can be seen by
inspection of Eqs.~(\ref{phiNA}) and (\ref{bNA}), a necessary
condition for this to happen in a finite interval in $\theta$ (to
ensure $\phi$ remains finite at all times) is that the curvature be
nonpositive. This is to be contrasted with the general relativistic
case for which a positive curvature is a pre-requisite to ensure that
a bounce is possible~\cite{ppnpn1}, and can be understood by stating
that, in the case at hand, a positive curvature implies a finite scale
factor at all times.

Under the assumption that both sides are positive definite, one can
integrate Eq.~(\ref{bNA}), written as
\begin{equation}
\int_{b_0}^{b} {\dd \tilde b \over \tilde b \sqrt{C^2 + M \tilde b^2 -
k \tilde b^4}} = \pm \int_{\theta_0}^\theta {\dd \tilde \theta\over
\phi_0 \sin(\alpha\tilde \theta)}, \label{integr}
\end{equation}
to provide the solution [see, e.g., Ref.~\cite{Grad}, Eq.~(2.266)]
\begin{equation}
{g(b)\over g(b_0)} = {f(\theta)\over f(\theta_0)},
\end{equation}
where
\begin{equation}
f(\theta) = \left| \tan
\left(\frac{\alpha\theta}{2}\right)\right|^p,\label{fNA}
\end{equation}
and
\begin{equation}
g(b) = {M\over C} + {2\over b^2} \left( C+\sqrt{C^2 + M b^2
-kb^4}\right),
\end{equation}
$b_0$ and $\theta_0$ being constants of integration that we choose
such that $C g(b_0) = f(\theta_0)$ for further convenience, and
\begin{equation}
p = \pm \sqrt{1 + \frac{2}{3}\omega}.
\end{equation}
Setting $a_0^2 = 4 C^2/\phi_0$, we finally get
\begin{equation}
a(\theta) =
\frac{a_0}{\sqrt{\sin\alpha\theta}}\left\{\frac{f(\theta)}
{\left[M - f(\theta)\right]^2 + 4C^2k}\right\}^{1/2}\label{ana},
\end{equation}
which is the desired result for the scale factor. Note that because
of the trigonometric identity
\begin{equation}
\tan \left[ -\left( {\alpha\theta +\pi/2\over 2}\right) \right] =
\left[ \tan \left( {\pi/2-\alpha\theta \over 2}\right) \right]^{-1},
\end{equation}
the solution~(\ref{ana}) with $p\to -p$ can be straightforwardly
deduced from the original one by a mirror symmetry with respect to the
point $\alpha\theta=\pi/2$. It is thus sufficient to consider $p>0$
and we shall in what follows restrict our attention to this case.

These solutions have some interesting features. As we have already
discussed, for $k = 1$, there are no bouncing solutions. On the other
hand, for $k = 0$ or $k= - 1$, it is possible to choose the parameters
in such a way that the extremes of the range of validity of the
variable $\theta$ occur for $t \rightarrow \pm \infty$, where
spacetime becomes flat.

The case $k=0$ was presented in Ref.~\cite{picco}; let us recall it
briefly for the sake of completeness.  The denominator in
Eq.~(\ref{ana}) has only two roots if $k=0$, and the parameter
$\theta$ varies from $\theta_{\rm i} =0$ to $\theta_{\rm f} =
2\alpha^{-1}\arctan(M^{1/p})$. Bouncing nonsingular solutions are
possible only when $-3/2 < \omega < -4/3$. This can be seen by
considering the limit for which $\theta \to \theta_{\rm i}=0$. There,
the scale factor is $a\propto\theta^{(p-1)/2}$, and, from
Eq.~(\ref{parameter}), $t\propto \theta^{(3p-1)/3}$, yielding
$a\propto |t|^{(p-1)/(3p-1)}$. As $a(t)$ is a power law (disregarding
the exceptional cases $p=1\Leftrightarrow\omega=0$, and
$p=1/3\Leftrightarrow\omega = -4/3$, also discussed in
Ref.\cite{picco}), the scalar curvature for $k=0$ is proportional to
$t^{-2}$, which converges (in fact, goes to zero) only if $t \to
\infty$ as $\theta \to 0$. This happens only for $p<1/3$, which yields
$-3/2 < \omega < -4/3$.  Note, however, that for $\theta=0$ the
dilaton $\phi$ vanishes, independently of the value of $\omega$,
rendering dubious the validity of the tree level action
(\ref{lagrange}) in this region.

When $k = -1$, the denominator in Eq.~(\ref{ana}) has now three
roots. One can take the parameter $\theta$ varying from $\theta_{\rm
i} =0$ to $\theta_{\rm f} =2\alpha^{-1}\arctan[(M-2C)^{1/p}]$,
supposing $2C < M$. In this interval, the same analysis given in the
precedent paragraph is valid here: one can have bouncing solutions
which present an initial singularity in the curvature and in the
string expansion parameter $g_{\rm s}^2 = \phi^{-1}$ if $-4/3< \omega
< 0$, and other bouncing solutions which do not present curvature
singularities initially but still have a singularity in the string
expansion parameter if $-3/2< \omega < -4/3$.  One can also take the
parameter $\theta$ to vary from $\theta_{\rm i}
=2\alpha^{-1}\arctan[(M-2C)^{1/p}]$ to $\theta_{\rm f} =
2\alpha^{-1}\arctan[(M+2C)^{1/p}]$. Provided $2C < M$, the dilatonic
field given by Eq.~(\ref{phiNA}) is finite and never vanishes, taking
constant values in the asymptotic regions.

Let us now consider the limit $\theta\to\theta_{\rm i}$ or
$\theta\to\theta_{\rm f}$. Setting $\alpha\theta = \alpha\theta_{\rm
i} +\varepsilon$ or $\alpha\theta = \alpha\theta_{\rm f}
-\varepsilon$, and expanding the denominator around $\varepsilon =0$,
we get $a\propto \varepsilon^{-1/2}$, from Eq.~(\ref{parameter})
$|t|\propto \varepsilon^{-1/2}$, and finally $a\propto |t|$,
independently of the value of $p$ or $\omega$.  As we are considering
$k=-1$, this limit corresponds to Milne flat spacetime. The scale
factor given by Eq.~(\ref{ana}) thus appears to represent, with this
choice of range for $\theta$, a universe contracting from a Milne
spacetime to a minimum size, bouncing to an expansion phase, and
ending asymptotically also in a Milne spacetime.

However interesting this solution might be, it is unfortunately
physically meaningless since it demands the dilaton to be
negative. This can be seen by looking at Eqs.~(\ref{bprim}) and
(\ref{ana}). From Eq.~(\ref{ana}) for $k=-1$, one can obtain the scale
factor in the Einstein frame $b(\theta)=\sqrt{\phi(\theta)}a(\theta)$,
which is
\begin{equation}
b(\theta) = 4C^2\left\{\frac{f(\theta)}
{\left[M - f(\theta)\right]^2 - 4C^2}\right\}^{1/2}\label{ana100}.
\end{equation}
In the above range of values of $\theta$, there must exist a point at
which $b'(\theta)=0$.  However, from Eq.~(\ref{bprim}) with $-3/2 <
\omega$, this could only be possible if $b(\theta)$ were purely
imaginary. Then, for the string-frame scale factor $a(\theta)$ to be
real, $\sqrt{\phi(\theta)}$ should also be imaginary, so that
$\phi(\theta)$ should be negative in the corresponding range of
values.
Many of the characteristic features of this solution are
however present also in the cosmologically more relevant anomalous
situation to which we now turn.

\subsection{Anomalous axionic case}

For $3 + 2\omega <0$, the previous solution for Eq.~(\ref{scalar'})
must be replaced by
\begin{equation}
\phi(\theta) = \phi_0\sinh (\alpha\theta),\label{phiAA}
\end{equation}
where now
\begin{equation}
\alpha = \sqrt{\frac{-2A^2}{3 + 2\omega}},
\end{equation}
and, as before, we have imposed $\phi(0)=0$.

Again, inserting this solution into Eq.~(\ref{bprim}) yields
\begin{equation}
\phi_0^2 \sinh^2(\alpha\theta) b'^2 = (M b^2 -C^2 - k b^4 ) b^2,
\label{bAA}
\end{equation}
where $C$ is as before [Eq.~(\ref{C})]. The same argument concerning
the existence of a bouncing solution applies, namely, that such
solutions cannot exist for $k=1$. Manipulations similar to those of
the previous case then lead to
\begin{eqnarray}
\label{fAA}
f(\theta) &=& \ln\left[\left| \tanh
\left(\frac{\alpha\theta}{2}\right) \right|^p \right],\\ g(b) &=&
\arcsin \left( {M b^2 - 2 C^2\over b^2 \sqrt{M^2-4kC^2} }\right) ,
\end{eqnarray}
where we have assumed $M^2-4kC^2 > 0$ (recall we are only interested
in the cases $k=0$ and $k=-1$). We now choose $f(\theta_0) = g(b_0)$,
set
\begin{equation}
p = \pm \sqrt{-\left( 1 + \frac{2}{3}\omega\right)}, \ \ \hbox{ and }
\ \ \ a_0^2 = {A^2 \phi_0\over 3 M},
\end{equation}
to obtain the scale factor as
\begin{equation}
\label{ana2}
a(\theta) = \frac{a_0}{\sqrt{\sinh\alpha\theta}}\left[1 \pm \sqrt{1 -
4\displaystyle{\frac{kC^2}{M^2}}}\sin f(\theta)\right]^{-1/2}.
\end{equation}
Regular bouncing solutions may be obtained for $k = 0$ or $k= - 1$.
Differently from the previous situation, the flat case also does not
exhibit any singularity in the string expansion
parameter. Furthermore, it is possible to find ranges of values of
$\theta$ for which the dilaton is strictly positive all along. This is
because Eq.~(\ref{bprim}) with $\omega <-3/2$ admits $b'(\theta)=0$
in the Einstein frame with a real scale factor $b$. Hence, it is not
necessary to have $\phi<0$ in order to have the string frame scale
factor $a$ real. As the dilaton is finite and strictly positive, there
are also no singularities in the string expansion parameter given by
$g_{\rm s}^2 = \phi^{-1}$, and the tree level approximation can be
trusted all along. Consequently, we have obtained a perfectly regular
bouncing solution in the string framework, without any singularity,
even in the dilatonic field, when the curvature of the spatial section
is negative or vanishing.

Investigation of the asymptotic behaviors reveals that, for $k = 0$,
the universe displays a radiation dominated phase in both extremities
of the range ($a\propto 1/|\theta|$, \ie $a\propto |t|^{1/2}$ for $t
\rightarrow \pm \infty$), while for $k = - 1$, the curvature dominates
in the asymptotic regions, leading to a Milne universe ($a\propto
|\theta|^{-1/2}\rightarrow |t|$). Hence, in the $k=0$ case, we have a
bounce between two asymptotic radiation dominated standard
cosmological models, one contracting and the other expanding, while
for $k=-1$ the bounce connects two Milne asymptotic regions.

Recovering the units and connecting the parameters with the real
Universe in order to evaluate the value of $M$, we consider $G_{\rm
eff} = \GN/\phi$, where $\GN$ is the value of the gravitational
coupling today, and make the replacements $\phi \rightarrow \phi/\GN$,
$\Psi \rightarrow \Psi/\GN$, $t\rightarrow a_0 t$, $a_0 \approx 1/H_0$
($H_0$ being the present Hubble parameter, which we choose to be our
inverse unit of time). Assuming the present amount of radiation ($\rho
_{0{\rm r}} = \Omega_{0{\rm r}}\rho_{\rm c}\sim 10^{-4} \rho_{\rm c}
\approx 10^{-33} {\rm g/cm}^3$, with $\rho_{\rm c}$ the critical
density today), we obtain from Eq.~(\ref{em1}), assuming the radiation
term to dominate at the time under consideration, that $M\sim 8\pi \GN
\rho_{0{\rm r}} H_0^{-2}/3 = \Omega_{0{\rm r}} \approx 10^{-4}$.

It is interesting to note that these models can provide a quite
effective way of enhancing the gravitational coupling. To illustrate
this point with a numerical example, let us choose $p = 1$ (\ie the
case derived in Sec.~\ref{sec:omega} with $\omega = - 3$, our
prototypical example), $k=0$, $f(\theta_{\rm i})=-7\pi/2$ and
$f(\theta_{\rm f})=-3\pi/2$. One then obtains $\phi_{\rm i}\approx
10^{-3}$ and $\phi_{\rm f}\approx 1$, where the constant $\phi_0$ is
chosen $\phi_0\approx 10^{-2}$ in order to obtain the effective
gravitational ``constant'' today equal to Newton constant $\GN$. With
this choice of parameters, the enhancement of the effective
gravitational ``constant'' in the past was therefore of three orders
of magnitude.  Note that the dilaton is strictly positive and finite
in this range.

\subsection{Normal RR case}

Integrating the equations of motion (\ref{scalar'}) after inclusion of
$\xi$, i.e., with a nonvanishing $B$ and still for $\omega > -3/2$,
simply turns the solution given by Eq.~(\ref{phiNA}) into
\begin{equation}
\phi(\theta) = \phi_0\left(\sin\alpha\theta - s\right),
\end{equation}
where
\begin{equation}
s = \frac{B^2}{2A^2\phi_0},
\end{equation}
provides the particular solution of the inhomogeneous equation, and we
have assumed the same initial condition for the homogeneous part. The
constant $\alpha$ is defined as in the normal axionic case.

After some straightforward calculations, we get that
Eq.~(\ref{integr}) is modified into
\begin{equation}
\int_{b_0}^{b} {\dd \tilde b \over \tilde b \sqrt{\pm C^2 + M \tilde
b^2 - k \tilde b^4}} = \pm \frac{1}{\phi_0}\int_{\theta_0}^\theta
\frac{\dd \tilde \theta}{\sin(\alpha\tilde\theta)-s},
\label{integr2}
\end{equation}
where now
\begin{equation}
C^2 = \frac{1}{6}A^2\phi_0^2 |1 - s^2| \label{Cs}
\end{equation}
takes into account the inhomogeneous part. In Eq.~(\ref{integr2}), the
sign in front of the factor $C^2$ in the denominator of the right-hand
side integrand is positive or negative depending on whether $s^2<1$ or
$s^2>1$ respectively. We shall treat both cases separately.

\subsubsection{Small RR-scalar}

We assume from now on that even though we allow variations for $\xi$,
those are limited in such a way that $s^2 <1$. Equation
(\ref{integr2}), being in a form similar to Eq.~(\ref{integr}), yields
the same result that bounces cannot be realized unless $k\leq 0$.

Integrating both sides of Eq.~(\ref{integr2}), we obtain the function
$b$, thanks to which we can write the scale factor as
\begin{equation}
a(\theta) = \frac{a_0}{\sqrt{\sin\alpha\theta -
s}}\left\{\frac{f(\theta)} {\left[ M - f(\theta)\right]^2 + 4C^2k }
\right\}^{1/2},\label{aNRR1}
\end{equation}
with [see again Ref.~\cite{Grad}, Eq.~(2.551/3)]
\begin{equation}
f(\theta) = \left| {2\over s}
\left[\frac{s\tan\left(\alpha\theta/2\right) - 1 + \sqrt{1 - s^2}}{1 +
\sqrt{1 -s^2}- s\tan\left(\alpha\theta/2\right)}\right]\right|^p,
\label{fNRR}
\end{equation}
where $a_0$, $p$ and the choice for the relationship between
$f(\theta_0)$ and $g(b_0)$ are the same as in the normal axionic case,
except for the new definition~(\ref{Cs}) of the constant $C$. The
normalization in Eq.~(\ref{fNRR}) has been chosen in such a way that
the limit $s\to 0$ gets indeed back to the normal axionic case
(\ref{fNA}).

The properties of these solutions are qualitatively the same as in the
normal axionic case. 

\subsubsection{Large RR-scalar}

In the opposite situation for which $s^2 >1$, one can normalize the
solution in such a way that [see Ref.~\cite{Grad}, Eq.~(2.551/3)]
\begin{equation}
f(\theta) = 2 p \arctan \left[
{1-s\tan\left(\alpha\theta/2\right)\over \sqrt{s^2-1} }\right],
\end{equation}
and, provided $a_0^2 = 2 C/(M\phi_0)$, the solution can be written as
\begin{equation}\label{aNRR2}
a(\theta) = \frac{a_0}{\sqrt{s - \sin(\alpha\theta)}}\left[1 \pm
\sqrt{1 - 4\displaystyle{\frac{kC^2}{M^2}}}\sin
f(\theta)\right]^{-1/2} \!\!\!\!\!\!\!\!\!\!.
\end{equation}

These solutions, both in the large and small RR-scalar sectors, share
with the normal axionic case the feature of requiring a meaningless
negative dilaton field. They were derived here for the sake of
completeness.

\subsection{Anomalous RR case}

Finally, the last situation, for which Eq.~(\ref{scalar'}) is solved
by
\begin{equation}
\phi =\phi_0 \left[\sinh\left(\alpha\theta\right)-s\right],
\end{equation}
is very similar to the anomalous axionic case except that the
hyperbolic sine squared in Eq.~(\ref{bAA}) is replaced by $[\sinh
(\alpha\theta) -s]^2$, with the same definition for the constant $s$
and $C$ as in the anomalous axionic situation. This case is
essentially similar to the normal RR one, except that we obtain a
different function [see Ref.~\cite{Grad}, Eq.~(2.441/3)]
\begin{equation}
f(\theta) = p\ln\left| {2\over s}
\left[\frac{s\tanh\left(\alpha\theta/2\right) + 1 - \sqrt{1 + s^2}}{1
+ \sqrt{1 + s^2}+ s\tanh\left(\alpha\theta/2\right)}\right]\right|,
\end{equation}
where the normalization again ensures that the limit $s\to 0$ is
equivalent to the anomalous axionic case. With the new scale factor
normalization
\begin{equation}
a_0^2 = {A^2 \phi_0\over 3M}(1+s^2),
\end{equation}
the new solution is expressed as
\begin{equation}
\label{aARR}
a(\theta) = \frac{a_0}{\sqrt{\sinh\alpha\theta - s}}\left[1 \pm
\sqrt{1 - 4\displaystyle{\frac{kC^2}{M^2}}}\sin
f(\theta)\right]^{-1/2} \!\!\!\!\!\!\!\!\!\!.
\end{equation}
Again, as in the anomalous axionic case, completely nonsingular
solutions, also with respect to the dilatonic field, which is strictly
positive, are obtained for
$k = 0$ or $k= - 1$. The properties of both anomalous (axionic and RR)
cases are very similar, even in the asymptotic regions.  The
significant feature of this case is that, for $k=0$, it is not
difficult to choose the free parameters in order to allow huge
increases of the dilaton along the evolution of such universes.

The anomalous cases are the ones that present bouncing eras connecting
asymptotically contracting and (standard) expanding cosmological
models which can represent the real Universe (with attractive
gravitation). The key requirement to obtain these solutions is that
$\omega < -3/2$, a property that can be obtained from $F$ theory in
twelve dimensions.

The presence of the axion and/or the RR scalar field is not important
qualitatively. They just change the functions $f(\theta)$ which appear
in the scale factor given by Eqs.~(\ref{ana}), (\ref{ana2}),
(\ref{aNRR1}), (\ref{aNRR2}) and (\ref{aARR}).  In fact, one can find
solutions without the axion, or with neither the axion nor the RR
scalar field, which are also given by the very same equations but with
different (and actually simpler) $f(\theta)$. For instance, in the
case where neither the axion nor the RR scalar field are present, we
have $[f(\theta)]^{1/p}=|\phi(\theta)|\propto |\theta|$ and
$[f(\theta)]^{1/p}=\ln(|\phi(\theta)|)=\ln(|\theta|)$ in
Eqs.~(\ref{ana}) and (\ref{ana2}), respectively. Those solutions can
also be obtained as limiting cases of Eqs.~(\ref{fNA}) and (\ref{fAA})
as $A\to 0$. The qualitative behavior of these solutions is the same
as in the case with the axion (normal and anomalous axionic cases).

\section{The null energy condition}

For a nonpositive curvature universe described by general relativity,
the null energy condition (NEC) $\rho_{_\mathrm{T}} +
p_{_\mathrm{T}}\geq 0$, where the subscript ``T'' denotes the total
contribution of all the fields and type of matter, must be violated in
order for a bounce to occur (see Ref.~\cite{Visser} and references
therein for a discussion of the relevant singularity theorems in
general relativity~\cite{singularity}). In the context under
consideration here, this result cannot be straightforwardly applied
since the nonminimal coupling involved prevents an easy
identification of the energy density and pressure sourcing the
Einstein geometry (many results can however be applied to a theory
with nonminimal coupling; this is discussed in, \eg
Ref.~\cite{singularity}).

\begin{figure}[t]
\begin{center}
\includegraphics[width=9cm]{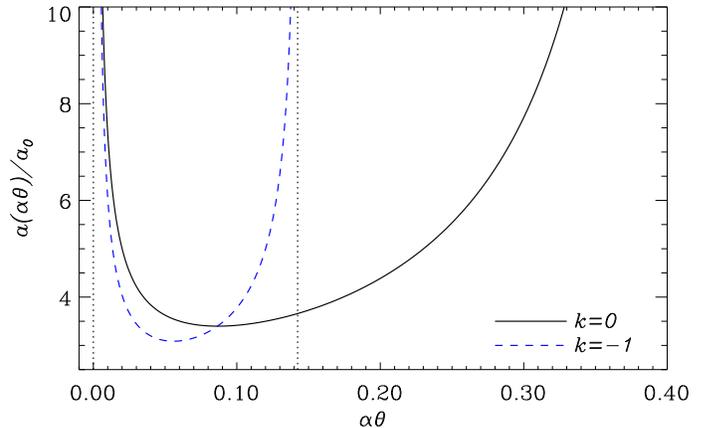}
\caption{Scale factors as functions of the parameter $\alpha\theta$
for two cases of interest, described by Eq.~(\ref{ana2}) in the case
derived in Sec.~\ref{sec:omega}, \ie with $\omega=-3$. The flat case
is shown as the full line, whereas the open case is plotted as a
dashed line. For these figures, the parameters have been chosen to be
$M=1.1$ and $C=1$. The straight dotted line at $\alpha\theta\sim 0.142$
represents the maximal permitted value for the timelike coordinate
$\alpha\theta$, \ie it corresponds to the future infinity, in the open
case $k=-1$, the equivalent line for the flat case $k=0$ being at the
edge of the figure (\ie the allowed range of variation is in this case
$0\leq \alpha\theta\lta 0.4$).}
\label{scalefactor}
\end{center}
\end{figure}

In what follows, we sketch an analysis of the problem of violation of
the energy conditions, adapting the expressions for the nonminimal
coupling with the dilatonic field. The effective energy density
$\rhoeff$ and pressure $\peff$ are derived from the field equations
(\ref{eq:Einstein}), whose right-hand side we take to be the effective
stress-energy tensor $T^{\mu\nu}_\mathrm{eff}$ we are looking for,
assuming an Einstein-like form for Eq.~(\ref{eq:Einstein}) as
\begin{equation}
R_{\mu\nu} - \frac{1}{2}g_{\mu\nu}R = 8\pi \GN
T_{\mu\nu}^\mathrm{eff}.
\label{Reff}
\end{equation}
The corresponding cosmological equations are then seen as projections
of Eq.~(\ref{Reff}) by means of a normalized timelike vector $u_\mu$
(with $u_\mu u^\mu =1$): defining $\rhoeff = T^{\mu\nu}_\mathrm{eff}
u_\mu u_\nu$ and $\peff = - \frac{1}{3}T^{\mu\nu}_\mathrm{eff} \left(
g_{\mu\nu} - u_\mu u_\nu\right)$, Eqs.~(\ref{eq:Einstein}) are then
nothing but the usual system describing a cosmological background with
a fluid.

Starting with the field equations derived above, one obtains the
following expressions:
\begin{eqnarray}
8\pi \GN \rhoeff &=& \frac{8\pi}{\phi}\rho +
\frac{\omega}{2}\frac{\dot\phi^2}{\phi^2} +
\frac{1}{2}\frac{\dot\Psi^2}{\phi^2} +
\frac{1}{2}\frac{\dot\xi^2}{\phi}\nonumber \\ & & - 3\frac{\dot
a}{a}\frac{\dot\phi}{\phi}, \label{rhoeff}\\ 8\pi \GN\peff &=&
\frac{8\pi}{3\phi}\rho + \frac{\omega}{2}\frac{\dot\phi^2}{\phi^2} +
\frac{1}{2}\frac{\dot\Psi^2}{\phi^2}
+\frac{1}{2}\frac{\dot\xi^2}{\phi} \nonumber \\ & & +
\frac{\ddot\phi}{\phi} + 2\frac{\dot
a}{a}\frac{\dot\phi}{\phi},\label{peff}
\end{eqnarray}
which, contrary to the standard cosmological situation, are both not
positive definite. The energy conditions may therefore be violated,
due to the terms arising from the nonminimal coupling and if either
$\omega$ or $\phi$ is negative.

To understand the causes for the bounces in our solutions, it is also
useful to use Eq.~(\ref{bprim}). In the normal cases (from the purely
mathematical point of view) with nonvanishing dilaton, we have seen
that one must have $\phi<0$ (repulsive gravity) in order to have a
bounce. In this case, the negative dilaton is sufficient to make the
bounce, and there is no need to have $\omega < 0$.  In the normal
cases with the dilaton vanishing at some point, $b$ does not actually
bounce. Hence the dilaton $\phi$ can be positive. The bounce then
comes from the requirement $\omega <0$.

\begin{figure}[t]
\begin{center}
\includegraphics[width=9cm]{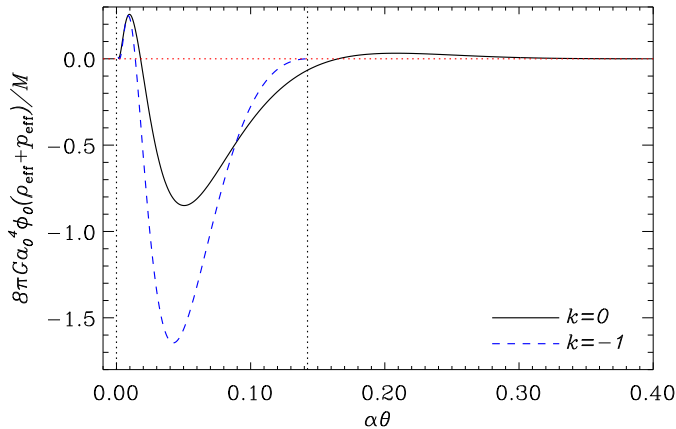}
\caption{The null energy condition during the evolution of the
Universe models depicted in Fig.~\ref{scalefactor}, again with the
full line standing for the $k=0$ case (rescaled by a factor of $3$ to
appear with comparable magnitude) and the dashed line for $k=-1$.}
\label{rho+p}
\end{center}
\end{figure}

In the anomalous cases, where the dilaton is strictly positive and
finite, it is the negative value of $\omega$, more precisely the fact
that $\omega <-3/2$, which causes the bounces. Figure
\ref{scalefactor} illustrates two cases of bounces for which the scale
factors are shown, taken in the anomalous axionic case, \ie using the
solution (\ref{ana2}). In terms of the scale factor, the expression
for the null energy condition reads
\begin{equation}
8\pi \GN \left(\rhoeff + \peff\right) = \frac{2}{a^6}
\left[ 4\left( \frac{a'}{a} \right)^2 - \frac{a''}{a} + k a^4\right],
\end{equation}
which coincides, on shell, with what is obtained by summing
Eqs.~(\ref{rhoeff}) and (\ref{peff}).  Plotting the right hand side of
this expression as in Fig.~\ref{rho+p} for the cases of
Fig.~\ref{scalefactor}, we verify that the null energy condition can
be violated in large domains around the bounce, depending on the
choice of parameters. One should notice that the timelike coordinate
$\theta$ has been chosen to emphasize the bounce itself. In terms of
the cosmic time, the domain where the NEC is violated is in fact
reduced to just a small fraction of the whole interval, since the
latter is actually infinite. Moreover, no violation is observed in the
large positive time limit, where the models tend to Milne or radiation
dominated universes.

\section{Conclusions}

We have constructed fully regular cosmological solutions in the
framework of effective actions derived from string theory
principles. These solutions present bouncing behaviors for a wide
range of parameters ($\omega < -3/2$), and are singularity free;
furthermore, the spacetimes they lead to are geodesically complete,
thereby improving the so-called horizon problem of standard
cosmology. Stemming from string theory in the context of the so-called
$F$ theory in twelve dimensions, where it is possible to have $\omega
< -3/2$, they have a reasonably sound basis as long as the dilaton is
strictly positive and finite in such a case. As a consequence, it is not
necessary to go beyond the tree level approximation in any part of
their histories: the analytic solutions exhibited above can describe
the whole history of the cosmological models they represent. Their
consequences may, in turn, be used as cosmological tests.

Remembering that the radiation fluid included here also has a
motivation in the superstring type IIB action, this turns out to be,
to our knowledge, the first case where a complete regular bouncing
cosmological solution is obtained in the string framework and related
theories, which moreover is smoothly connected with the standard
cosmological model radiation dominated phase. This solution may have
flat or negative curvature spatial sections.

In the axionic and RR cases with $k=0$ or $k=-1$, there are
nonsingular bouncing solutions for $-3/2<\omega<-4/3$ but with
vanishing dilaton in the beginning, where the tree level action cannot
be trusted, and bouncing solutions with an initial curvature
singularity if $-4/3<\omega<0$. If $k = - 1$ and $-3/2<\omega<0$
(normal, including the pure string case), one can have singularity
free bouncing solutions with, however, a negative definite dilaton
field.

As all the models with $\omega<-3/2$ have the interesting feature to
approach flat spacetime in the infinite past (either in Milne
coordinates for $k=-1$, or the infinitely large radiation dominated
standard model with $k=0$), there is the possibility to implement a
quantum spectrum of perturbations in the initial asymptotics without
any trans-Planckian problem~\cite{transP}, and, at the same time, to
accomplish a smooth transition to the standard cosmological model
when, after the bounce, a standard radiation dominated phase is
recovered (asymptotically in the $k=0$ case), preserving some of its
main achievements like, \eg primordial nucleosynthesis. The bouncing
solutions with $\omega>-3/2$ and positive dilaton still present some
sort of trans-Planckian problem as long as the string expansion
parameter diverges initially and one must go beyond the tree level
action in such cases.

Notice that, in the cases where the dilaton is strictly positive, the
initial value of the dilatonic field can be made smaller than its
final value. Hence, the gravitational coupling can be initially given
a much greater value than it would have today. This opens the
possibility to solve the hierarchical problem of the gravitational
coupling, in a spirit similar to the so-called brane
cosmology~\cite{langlois}.

\acknowledgments We thank CNPq (Brazil) for financial support.

\end{document}